\documentstyle[12pt,a41,epsfig]{article}

\newcommand{\ds}{\displaystyle}
\newcommand{\bq}{\begin{equation}}
\newcommand{\eq}{\end{equation}}

\newcommand{\gsim}{\raisebox{-0.07cm}{$\,\stackrel{>}{{\scriptstyle
 \sim}}\, $} }
\newcommand{\lsim}{\raisebox{-0.07cm}{$\,\stackrel{<}{{\scriptstyle
 \sim}}\, $} }

\newcommand\MeV{\,\mbox{MeV}}
\newcommand\GeV{\,\mbox{GeV}}

\newcommand\GA{\,\mbox{\boldmath $\Gamma$}}

\begin{document}
\setlength{\baselineskip}{0.525cm}
\sloppy
\thispagestyle{empty}
\begin{flushleft}
DESY 96--041 \\
March 1996\\
\end{flushleft}

\setcounter{page}{0}

\mbox{}
\vspace*{\fill}
\begin{center}
{\Large\bf On the Resummation of $\alpha \ln ^2 x$~Terms for} \\
\vspace{3mm}
{\Large\bf  Non--Singlet Structure Functions in QED and 
QCD{\Large\footnote{Invited
talk at the 1996 Cracow Epiphany Conference on Proton Structure,
5--6 January 1996, Krak\'ow, Poland, presented by J.~Bl\"umlein
}}}
\\
\vspace{5em}
\large
J. Bl\"umlein$^a$ and A. Vogt$^{b,}$\footnote{On leave of absence
from Sektion Physik, Universit\"at M\"unchen, D-80333 Munich, Germany}
\\
\vspace{5em}
\normalsize
{\it   $^a$DESY--Zeuthen}\\
{\it   Platanenallee 6, D--15735 Zeuthen, Germany}\\
\vspace{5mm}
{\it   $^b$Deutsches Elektronen-Synchrotron DESY}\\
{\it   Notkestra\ss{}e 85, D--22603 Hamburg, Germany}\\
\end{center}
\vspace*{\fill}
\begin{abstract}
\noindent
The resummation of $O(\alpha ^{l+1} \ln^{2l} x)$ terms in the evolution
kernels of non--singlet combinations of structure functions is 
investigated for both QED and QCD.
Numerical results are presented for unpolarized and polarized QCD
structure functions.
\end{abstract}
\vspace*{\fill}
\newpage
%
%
\section{Introduction}
\label{sect1}
%
%
The resummation of leading small-$x$ contributions to the evolution 
kernels of QCD singlet structure functions \cite{BFKL} may lead to large
effects~\cite{CCH}. The small-$x$ behaviour of the corresponding
anomalous dimensions is dominated by the leading singularity in the 
$N$-moment plane $\sim (\alpha_s/[N-1])^l$ and corrections to it. 
For non--singlet kernels such terms are absent both for unpolarized
and polarized deep-inelastic scattering, and the most singular 
contributions behave like $N(\alpha_s/N^2)^l$. A resummation of these 
terms was derived for QCD amplitudes in ref.~\cite{KL} more than a 
decade ago.

Similar terms emerge also in QED. There the resummed form of these 
contributions may be described by the structure function method. In 
the present paper, after setting up our notation and recalling the 
standard NLO formulation in section 2, we present in section 3 a 
derivation of the resummed kernels for the case of QED and QCD 
\cite{BV} in parallel to relate both cases in a direct way. 
The asymptotic kernels are compared with those results found in 
complete calculations in the limit $x \rightarrow 0$ up to 
next-to-leading order (NLO) in QED and QCD.

Recently very sizeable corrections due to this resummation have been
claimed~\cite{BEMR} for the structure functions in both unpolarized and 
polarized deep inelastic scattering at small $x$. In this way the 
small-$x$ behaviour of the structure function evolution, as e.g. for
$xF_3^{\,\nu d}(x,Q^2), F_2^{\, p}(x,Q^2) - F^{\, n}_2(x,Q^2)$, and 
$g_1^{\, p}(x,Q^2) - g_1^{\, n}(x,Q^2)$ should be considerably affected.
In section 4, we perform a detailed numerical analysis and derive the 
corrections to the various QCD non--singlet combinations\footnote{A part
of the results has been published in ref.~\cite{BV} recently.}.

Also in the case of the singlet anomalous dimensions for polarized deep 
inelastic scattering the leading singularity is expected to behave like 
$\sim  N(\alpha_s/N^2)^l$. Corresponding equations for related QCD
amplitudes have been given in~\cite{BER} recently.
The explicit form of the resummed anomalous dimension matrix as a 
function of $\alpha_s$ is derived in~\cite{BV1} where also numerical 
results on the behaviour of the structure function $g_1(x,Q^2)$ are 
presented. In the present paper we will deal with the different 
non--singlet cases only and refer for the singlet evolution to 
ref.~\cite{BV1}.
%
%
\section{Evolution in fixed--order perturbative QED and QCD}
\label{sect2}
%
%
The  evolution equation for the non--singlet combinations 
$q_{\rm NS}^{\pm}(x,Q^2)$ of parton densities is given by
\begin{equation}
\label{ev1}
 \frac{\partial q_{\rm NS}^{\pm}(x,Q^2)}{\partial \ln  Q^2}
 = P_{\rm NS}^{\pm}(x, \alpha) \otimes q_{\rm NS}^{\pm}(x,Q^2) \: .
\end{equation}
The corresponding splitting function combinations $P_{\rm NS}^{\pm}
(x,\alpha)$ are specified below, and $\otimes$ stands for the Mellin 
convolution. $\alpha$ denotes the running coupling constant in either 
QED or QCD. In order to simplify the notation, we will use the 
abbreviation $a \equiv \alpha/(4 \pi)$ in the following. The scale 
dependence of the running coupling is defined by
\begin{equation}
 \label{arun}
 \frac{da}{d\ln Q^2} = - \sum_{k=0}^{\infty} a^{k+2}\beta_k \: .
\end{equation}
The first two coefficients of the $\beta$--function, $\beta_0$ and 
$\beta_1$, are independent of the renormalization scheme. For (one 
flavour) QED and QCD they read  
\begin{equation}
\begin{array}{lcllcl}
\beta_0^{\,\rm QED} &=& - \frac{\ds 4}{\ds 3} \: ,
 & \beta_1^{\,\rm QED} & = &  -4 \: , \\
 & & & & & \\
\beta_0^{\,\rm QCD} &=& \frac{\ds 11}{\ds 3} C_G - \frac{\ds 4}{\ds 3} 
 T_R N_f \: ,
 & \beta_1^{\,\rm QCD} & = & \frac{\ds 34}{\ds 3} C_G^2 - \frac{\ds 20}
 {\ds 3} C_G T_R N_f -  4 C_F T_R N_f \: ,
\end{array}
\end{equation}
with $C_G = N_c \equiv 3, C_F = (N_c^2-1)/(2 N_c) \equiv 4/3$, $T_R = 
1/2$, and $N_f$ the number of flavours in the QCD case.

In what follows we drop the subscript NS wherever the non--singlet 
character of the considered quantity is obvious from the superscript 
$\pm$. The splitting functions $P^{\pm}(x,a)$ are given by the 
combinations
\begin{equation}
 P^{\pm}(x,a) = P_{qq}(x,a) \pm P_{q\overline{q}}(x,\alpha) \equiv
 \sum_{l=0}^{\infty} a^{l+1} P_l^{\pm}(x) \: .
\end{equation}
For the subsequent analysis we restrict ourselves to the consideration 
of the spacelike case $Q^2 = -q^2 > 0$. Due to fermion number 
conservation, the expansion coefficients $P^-_l$ are subject to the 
sum rule  
\begin{equation}
 \int_0^1 \! dx \, P^-_l(x) = 0 \: ,
\end{equation}
since $a$ acts as an independent parameter.
The non--singlet splitting functions for QCD are known up to NLO 
\cite{NS1,NS2} and read in the $\overline{\rm MS}$ factorization scheme
\begin{eqnarray}
\label{PNLO}
 P_{qq}(x, a) &=& 2 a C_F \left[ \frac{1 + x^2}{1 - x} \right]_+
   \nonumber \\
 & & \mbox{} + a^2 \left [ C_F^2 P_F(x) + \frac{1}{2} C_F C_G P_G(x) +
    C_F N_f T_R P_{N_f}(x) \right ] + {\cal O}(a^3) \: ,    \\
 P_{q\overline{q}}(x, a) &=& 4 a^2 \left [ C_F^2
   - \frac{1}{2} C_F C_G \right ] P_A(x) + {\cal O}(a^3) \: .
\label{PNLO1}
\end{eqnarray}
The functions $ P_I (x),\, I = F,\, G,\, N_f,\,$ and $ A $, can be 
found in refs.~\cite{NS2}.
The corresponding splitting functions for QED in this scheme are 
obtained from (\ref{PNLO}) and (\ref{PNLO1}) by setting $C_F = T_R = 1$
and $C_G = 0$. Most QED calculations are carried out, however, in the 
on--mass-shell (OMS) scheme for which the NLO splitting functions are 
different \cite{BBN}.
For $x \rightarrow 0$ the leading contributions to $P^{\pm}(x, a)$ in
the $\overline{\rm MS}$ factorization scheme are 
\begin{eqnarray}
 P^{+, \rm
QED}_{x \rightarrow 0} (x,a) &=& 2a + 2a^2 \ln^2 x + {\cal O}(a^3)
  \nonumber \\
 P^{-, \rm
QED}_{x \rightarrow 0} (x,a) &=& 2a - 6a^2 \ln^2 x + {\cal O}(a^3)
  \: ,  \label{eqB1} \\
 P^{+, \rm
QCD}_{x \rightarrow 0} (x,a) &=& 2a C_F + 2a^2 C_F^2 \ln^2 x
  + {\cal O}(a^3)  \nonumber \\
 P^{-, \rm
QCD}_{x \rightarrow 0} (x,a) &=& 2a C_F + 2a^2 \left [-3 C_F^2 
  + 2 C_F C_G \right ] \ln^2 x + {\cal O}(a^3) \: . \label{eqB}
\end{eqnarray}

Since the parton densities $q^{\pm}(x,Q^2)$ are scheme dependent and
hence no observables beyond leading order, it is convenient to consider 
also the evolution equations for the related observables directly. 
These are given by the corresponding structure functions $F_i^{\pm} 
(x,Q^2)$, obtained by the convolution
\begin{equation}
 \label{ev2}
 F_i^{\pm}(x,Q^2) = c_i^{\pm}(x,Q^2) \otimes q^{\pm}_i(x,Q^2) \: .
\end{equation}
Here $c_i^{\pm}(x,Q^2)$ denote the respective coefficient functions
which can be expanded in $a$ as
\begin{equation}
\label{coef}
 c_i^{\pm}(x,Q^2) = \delta(1 -x) + \sum_{l=1}^{\infty}
 a^l  c_{i,l}^{\pm}(x) \: .
\end{equation}
After transformation to an equation in $a$ using (\ref{arun}), the 
evolution equation for $F^{\pm}_i(x,Q^2)$ resulting from (\ref{ev1})
and (\ref{ev2}) reads
\begin{equation}
\label{EV}
 \frac{\partial F_i^{\pm}(x, a)}{\partial a} = - \frac{1}{\beta_0 a^2} 
  K_i^{\pm}(x, a) \otimes F_i^{\pm}(x, a) \: ,
\end{equation}
where in NLO the kernels can be written as
\begin{equation}
\label{KNS}
 K_{i,1}^{\pm}(x, a) = a\, P_{\rm NS,0}(x) + a^2 \,\left[ P_1^{\pm}(x)- 
 \frac{\beta_1}{\beta_0} P_{\rm NS,0}(x) - \beta_0 c_{i,1}^{\pm}(x) 
 \right] \: .
\end{equation}
The terms $\propto a (a \ln^2 x)^k $ emerge in the $a$-expansion of the 
kernels $ K_{i}^{\pm}(x, a) $ only in combination with the coefficient
$\beta_0$. In this sense the resummation to which we turn now is of 
leading order.
%
%
\section{Resummation of dominant terms in the limit $x \rightarrow 0$}
\label{sect3}
%
%
The most singular contributions to the Mellin transforms of the 
structure--function evolution kernels $K^{\pm}(x,a)$ at all orders in 
$a$ can be obtained from the positive and negative signature amplitudes 
$f_0^{\pm}(N, a)$ studied in ref.~\cite{KL} for QCD via
\begin{equation}
{\cal M} \left [ K_{x \rightarrow 0}^{\pm}(a) \right ](N) 
 \equiv \int_0^1 \! dx \, x^{N-1} K_{x \rightarrow 0}^{\pm}(x, a)
 \equiv - \frac{1}{2} \GA_{x \rightarrow 0}^{\pm}(N, a)
 = \frac{1}{8\pi ^2} f_0^{\pm}(N, a) \: .
\end{equation}
These amplitudes are subject to the quadratic equations:
\begin{eqnarray}
f_0^+(N, a) &=&  16 \pi^2 a_0  \frac{a}{N} + \frac{1}{8 \pi^2}
 \frac{1}{N} \left[ f_0^+(N, a) \right] ^2 \: , \label{eqf0p} \\
f_0^-(N, a) &=&  16 \pi^2 a_0  \frac{a}{N} + 8 b_0^-  \frac{a}{N^2}
 f_V^+(N, a) + \frac{1}{8 \pi^2} \frac{1}{N} \left [ f_0^-(N, a)
 \right ]^2 \: .  \label{eqf0m}
\end{eqnarray}
Here $f_V^+(N, a)$ is obtained as the solution of the Riccati 
differential equation
\begin{equation}
\label{eqRIC}
 f_V^+(N, a) = 16 \pi^2 a_V  \frac{a}{N} + 2 b_V  \frac{a}{N}
 \frac{d}{d N} f_V^+(N, a) + \frac{1}{8 \pi^2} \frac{1}{N}
 \left [f_V^+(N, a)\right]^2 \:\: .
\end{equation}
The coefficients $a_i$ and $b_i$ in the above relations read for the 
case of QED
\begin{equation}
 a_0 = 1,~~b_0^- = 1,~~a_V = 1,~~b_V = 0,
\end{equation}
and for QCD~\cite{KL}
\begin{equation}
 a_0 = C_F,~~b_0^- = C_F,~~a_V = -\frac{\ds 1}{2 N_c},~~b_V 
 = N_c.
\end{equation}
In QED eq.~(\ref{eqRIC}) further simplifies to an algebraic equation 
with the same coefficients as (\ref{eqf0p}). The solutions of
(\ref{eqf0p}) and (\ref{eqf0m}) were derived in ref.~\cite{KL} for the 
QCD case\footnote{Note that there are a few misprints in eq.~(4.7) of 
ref.~\cite{KL}.}. They are given by 
\begin{eqnarray}
 \GA_{x \rightarrow 0}^{+, \rm QED}(N, a)  &=& - N
  \left \{ 1 - \sqrt{1 - \frac{8 a}{N^2}} \right \} \nonumber \\
 \GA_{x \rightarrow 0}^{-, \rm QED}(N, a)  &=& - N
  \left \{ 1 - \sqrt{1 + \frac{8 a}{N^2} \left [ 1 -
  2 \sqrt{1 - \frac{8 a}{N^2}}~\right ] } \right \} \: , 
\label{eqGAA1}
\\
 \GA_{x \rightarrow 0}^{+, \rm QCD}(N, a) &=& - N
  \left \{ 1 - \sqrt{1 - \frac{8 a C_F}{N^2}} \right \} \nonumber \\
 \GA_{x \rightarrow 0}^{-, \rm QCD}(N, a) &=& - N
  \left \{ 1 - \sqrt{1 - \frac{8 a C_F}{N^2}
  \left [1 - \frac{8 a N_c}{N} \frac{d}{d N}
  \ln \left ( e^{z^2/4} D_{-1/[2N_c^2]}(z) \right ) \right ] } 
  \right \} \: , 
\label{eqGAA}
\end{eqnarray}
where $z = N/\sqrt{2 N_c a}$, and $D_p(z)$ denotes the function of the 
parabolic cylinder~\cite{RYGRA}.

It is instructive to expand the resummed anomalous dimensions 
(\ref{eqGAA1}) and (\ref{eqGAA}) into a series in $a^k/N^{2k-1}$ and 
transform the result to $x$--space using
\begin{equation}
\label{LTRF}
 {\cal M} \left [ \ln^k \left (\frac{1}{x} \right) \right ] (N) =
 \frac{k!}{N^{k+1}} \:\: .
\end{equation}
This results in 
\begin{eqnarray}
 K_{x \rightarrow 0}^{+, \rm QED}(x, a)  &=& 2 a + 2 a^2 \ln^2 x
   + \frac{2}{3} a^3 \ln^4 x + {\cal O}(a^4 \ln^6 x)
 \nonumber \\
 K_{x \rightarrow 0}^{-, \rm QED}(x, a)  &=& 2 a - 6 a^2 \ln^2 x
   - \frac{10}{3} a^3 \ln^4 x + {\cal O}(a^4 \ln^6 x) \: ,
 \label{eqPPN} \\
 K_{x \rightarrow 0}^{+, \rm QCD}(x, a) &=& 2 a C_F + 2a^2 C_F^2 \ln^2 x
   + \frac{2}{3} a^3 C_F^3 \ln^4 x + {\cal O}(a^4 \ln^6 x)
 \nonumber \\
 K_{x \rightarrow 0}^{-, \rm QCD}(x, a) &=& 2 a C_F + 2 a^2 C_F 
   \left [C_F + \frac{2}{N_c} \right ] \ln^2 x + \frac{2}{3} a^3 C_F 
   \left [ C_F^2 - \frac{3}{2 N_c} \right ] \ln^4 x
   \nonumber \\
   & & \mbox{} + {\cal O}(a^4 \ln^6 x) \: .
 \label{eqPPM}
\end{eqnarray}
Eqs.~(\ref{eqPPN}--\ref{eqPPM}) agree with the corresponding result 
found for $P_{NS, x \rightarrow 0}^{\pm}(x, a)$ in (\ref{eqB1}--\ref
{eqB}) in the complete NLO calculations of the non--singlet anomalous 
dimensions~\cite{NS2} in the most singular terms since
\begin{equation}
 C_G - \frac{3}{2} C_F = \frac{1}{N_c} + \frac{1}{2} C_F
\end{equation}
holds in $SU(N_c)$.
Moreover, $K_{x \rightarrow 0}^{-, \rm QED}(x, a)$ can be compared 
directly with a result in ref.~\cite{BBN}, eqs.~(2.30, 2.40, 2.43),
restricting to the terms $\propto a^2 \ln^2 x$, where the 
`--'-non--singlet\footnote{Note that the singlet contributions contain 
terms $\propto 1/x$ also in the case of QED.} terms were given 
separately in the OMS scheme. Since the corrections there refer to the 
initial state radiation in $e^+e^-$~annihilation the NLO result for a 
single (massless) fermion line reads
\begin{eqnarray}
\left.
K_{1,x \rightarrow 0}^{-, \rm QED}\right|_{\rm OMS}(x, a) &=&
 \frac{1}{2} \left( \frac{\alpha}{\pi} \right )^2 \left [
 \delta^{\rm I}_{e^+e^-} + \delta^{\rm II}_{e^+e^-} +
 \delta^{\rm IV}_{e^+e^-} \right ] \nonumber\\
&\equiv& \frac{1}{2} \left (\frac{\alpha}{\pi} \right )^2 \left [
 - \frac{1}{4} + 0 - \frac{1}{2} \right ] \ln^2 x = - 6 a^2 \ln^2 x
 = \left.
 K_{1,x \rightarrow 0}^{-,\rm QED}\right|_{\overline{\rm MS}}(x,a) 
 \:\: .
\end{eqnarray}
The corresponding result for $K_{x \rightarrow 0}^{+, \rm QED}$ can not 
be derived directly.

In the evolution equation~(\ref{ev2}) aside from the anomalous 
dimensions $P_{l}^{\pm}(x)$ also the coefficient functions 
$c_{i,l}^{\pm}(x)$ contribute. The latter quantities have been 
calculated to $O(a^2)$ (i.e.\ $l$=2)~\cite{CO1}--\cite{CO3} for 
$ xF_3(x,Q^2)$ and the non--singlet part of the structure functions 
$F_2(x,Q^2)$ and $g_1(x,Q^2)$ in the $\overline{\rm MS}$ scheme. 
Expanding these coefficient functions for $x \rightarrow 0$ one finds 
after noticing that apparent terms $\propto 1/x^m, m=1,2$ cancel in the
corresponding expressions of ref.~\cite{CO2,CO3}
\begin{equation}
 c_{i,1} \propto  \alpha_s \ln \left (\frac{1}{x} \right) \: , \:\:\:
 c_{i,2} \propto  \alpha_s^2 \ln^3 \left (\frac{1}{x} \right) \: .
\end{equation}
Hence the terms of $O(a^2)$ and $O(a^3)$ in (\ref{eqPPN}--\ref{eqPPM}) 
can be identified with the parts of the $\overline{\rm MS}$ non--singlet
splitting functions $\propto a(a \ln^2 x)^l$. Thus one obtains
\begin{eqnarray}
P_{2, x \rightarrow 0, {\overline{\rm MS}}} ^{+, \rm QED}(x, a)  &=& 
 \frac{2}{3} a^3 \ln^4 x  \nonumber \\
P_{2, x \rightarrow 0, {\overline{\rm MS}}} ^{-, \rm QED}(x, a)  &=& 
 - \frac{10}{3} a^3 \ln^4 x \: , \\
P_{2, x \rightarrow 0, {\overline{\rm MS}}} ^{+, \rm QCD}(x, a)  &=&
 \frac{2}{3} C_F^3 a^3 \ln^4 x  \nonumber \\
P_{2, x \rightarrow 0, {\overline{\rm MS}}} ^{-, \rm QCD}(x, a)  &=& 
 \left ( - \frac{10}{3} C_F^3 + 4 C_F^2 C_G - C_F C_G^2 \right )
 a^3 \ln^4 x \: .
\label{eqPPMX}
\end{eqnarray}
It should be noted that the agreement of the NLO terms between
(\ref{eqPPN}--\ref{eqPPM}) obtained from the above resummation and
(\ref{eqB1}--\ref{eqB}) holds for $q^2 < 0$ only. This is due to
the violation of the Gribov--Lipatov relation in the $ \ln^2 x$ term
of the NLO splitting functions for $q^2 > 0$.
%
%
\section{Numerical results for nucleon structure functions}
%
%
We now write down the solution of the evolution equation derived in 
the previous sections and then study the quantitative consequences of
the leading small-$x$ resummation. In the following we will confine 
ourselves to the QCD case.
Here we make use of the fact that the evolution equation (\ref{EV}) for 
non--singlet structure function combinations reduces to a single 
ordinary differential equation after transformation to Mellin moments. 
Including the effect of the resummed kernels (\ref{eqGAA}), the 
corresponding solution can be written as 
\begin{eqnarray}
\label{SOL}
  F^{\pm }(N,a_s) & = & F^{\pm}(N,a_0)
   \left( \frac{a_s}{a_0} \right)^{\gamma_{{\,\rm NS},0}(N)/2\beta_0} \\
   &\times&\! \left \{ \exp \left[ \frac{1}{2\beta_0} \int_{a_{0}}^{a_s}
     \! da \, \frac{1}{a^2} \Gamma ^{\pm}(N,a_s) \right]
   + \frac{a_s - a_0}{2\beta_{0}} \left[ \tilde{\gamma}_1^{\pm }(N) -
     \frac{\beta_1}{\beta_{0}} \gamma_{{\rm NS},0}(N) + 2\beta_0 
     \hat{c}_{i,1}(N)
 \right] \right \} \nonumber
\end{eqnarray} 
with
\begin{equation}
\gamma_i^{\pm}(N)
 = -2 \int_0^1 \! dx \, x^{N-1} P_i^{\pm}(x) \: , \:\:\:\:\:
\hat{c}_ i^{\pm}(N)
 =    \int_0^1 \! dx \, x^{N-1} c_i^{\pm}(x) \: ,
\end{equation}
and $a_0 = a_s(Q^2_0)$.
In (\ref{SOL}) $ \tilde{\gamma}_1^{\pm }(N) $ stands for the two--loop 
anomalous dimension $ \gamma_1^{\pm }(N) $ with the leading $ 1/N^3 $ 
term obvious from (\ref{eqB}) removed\footnote{In (\ref{SOL}) and 
(\ref{SOL1}) we have corrected two trivial misprints in eqs.~(21)
and (23) of ref.~\cite{BV}.}. This latter contribution is already 
included in the exponential factor, which in turn is connected to 
(\ref{eqGAA}) via the subtraction of the contribution linear in $a_s$,  
\begin{equation}
\label{SOL1}
  \Gamma ^{\pm}(N,a_s) = \GA _{x \rightarrow 0}^{\pm}(N,a_s) -
  \frac{a_s}{N} \lim_{N \rightarrow 0} \, [N\gamma_{{\,\rm NS},0}(N)] 
  = \GA _{x \rightarrow 0}^{\pm}(N,a_s) + a_s \frac{4C_F}{N} \: .
\end{equation}
The well--known NLO evolution of $F^{\pm}(N,a_s)$ is entailed in (\ref
{SOL}) by simply expanding the exponential to first order in $a_s$ and 
$a_0$. Finally, the transformation of the solution back to $x$--space
at any $x$ and $Q^2$ affords only one standard numerical integral in 
the complex $N$-plane \cite{GRV90}.

The remaining quadrature in (\ref{SOL}) can be performed analytically
for the `+'-case, resulting in
\begin{eqnarray}
 \int_{a_{0}}^{a_s} \! da \, \frac{1}{a^2} \Gamma ^{+}(N,a_s) & = &
   \frac{NA}{2} \ln \frac{a_s}{a_0} 
 + N \left( \frac{1}{a_s} - \frac{1}{a_0} \right)
 - N \left\{ \frac{\sqrt{1-Aa_s}}{a_s} - \frac{\sqrt{1-Aa_0}}{a_0} 
     \right\}   \nonumber \\
 & & \mbox{} \hspace*{-0.3cm}
 - \frac{NA}{2} \ln \frac{(1-\sqrt{1-Aa_s}) (1+\sqrt{1-Aa_0}) }
 {(1+\sqrt{1-Aa_s}) (1-\sqrt{1-Aa_0}) }  
 \:\: , \:\:\:\: A = \frac{8C_F}{N^2} \:\: .
\end{eqnarray}
On the other hand, the corresponding integration has been carried out
numerically for the `--'-combinations involving the parabolic cylinder
function $ D_p(z\! =\! N/\sqrt{2N_{c}a_s}) $. Alternatively, one can 
expand the resummed kernels $ \Gamma ^{+}(N,a_s) $ and $ \Gamma ^{-}
(N,a_s) $ in the strong coupling $a_s$, using the Taylor series of the 
square root and the asymptotic expansion\footnote{For $x \lsim 10^{-7}$
the asymptotic series is no longer reliable and one has to refer to the 
resummed result directly.} of $D_{p}(z)$, respectively. 
In the practical applications considered below, one finds that, even at 
the lowest $x$-values considered, more than 90\% of the resummation 
effects in (\ref{SOL}) arise from the first two terms beyond NLO in the 
$\alpha_{s}$ expansion of $ \Gamma ^{\pm}(N,a_s) $.

Let us consider the quantitative consequences of the resummation
(\ref{eqGAA}) for two representative non-singlet combinations. For this
purpose, we choose $ Q_{0}^{2} = 4 \mbox{ GeV}^2 $ as our reference 
scale in (\ref{SOL}), and employ the same initial distributions 
$F^{\pm}(N,a_0)$ and $\Lambda_{QCD}$ for the NLO and the resummed 
calculations. The evolution is performed for $N_f = 4$ active (massless)
quark flavours. Unless another value is stated explicitly, we take 
$ \Lambda \equiv \Lambda_{\overline {\rm MS}} (N_f = 4) = 230 \MeV$ in 
\begin{equation}
 a_s(Q^2) = \frac{1}{\beta_0 \ln(Q^2/\Lambda^2)} \left [ 1 - \frac{ 
 \beta_1}{\beta_0^2} \frac{\ln \ln(Q^2/\Lambda^2)} {\ln(Q^2/ \Lambda^2)}
 \right ] \: .
\end{equation}
 
We start with the unpolarized case, where we investigate the evolution 
of the `+'-combination\footnote{For corresponding results on the 
`--'-combination $xF_3 (x,Q^2)$ the reader is referred to 
ref.~\cite{BV}.}
\begin{equation}
\label{SOL2}
  F_{2}^{\, ep}(x,Q_0^2) - F_{2}^{\, en}(x,Q_0^2)
  = c_{F_2}^+(x,Q^2_0) \otimes \frac{1}{3}
  \left[ xu_v - xd_v - 2(x\bar{d}-x\bar{u})\right](x,Q^2_0) \: ,
\end{equation}
adopting the input densities from the MRS(A) \cite{MRSA} global fit. 
The small-$x$ behaviour of the most relevant quantities is given by 
$ xu_v(x,Q_{0}^{2}) \sim x^{0.54} $, $ xd_v(x,Q_{0}^{2}) \sim x^{0.33}$.
Note that these distributions are rather `steep', i.e.\ their rightmost
singularity in the complex $N$-plane lies about 0.5 units or more to the
right of the leading singularity of the non-singlet splitting functions
at $ N=0$.
Studying the evolution of this $F_2$ difference at very small $x$ is 
mainly of theoretical interest, since it is orders of magnitude 
smaller than $F_2^{\, ep}$ and $F_{2}^{\, en}$ there. In figure 1 the
result of this investigation is depicted down to $x$ as low as
$ 10^{-15} $. Even at these extremely low values of $x$, the effect of
the resummed anomalous dimensions stays at the level of 1\% or below,
and is still dominated entirely by the first two $\alpha_s$ terms 
beyond NLO.

In the polarized case we consider the corresponding difference
\begin{equation}
  g_{1}^{\, ep}(x,Q^2_0) - g_{1}^{\, en}(x,Q^2_0)
  = c^-_{g_1}(x,Q^2_0) \otimes \frac{1}{6}
  \left(\Delta u_v - \Delta d_v \right)(x,Q^2_0) \:\: .
\end{equation}
This case is practically much more interesting, firstly since -- unlike 
in the unpolarized case -- the non-singlet distributions are not a 
priori suppressed here with respect to the singlet ones at very low $x$,
and secondly since the shapes of the polarized initial distributions 
are not yet well established. We illustrate the strong dependence of the
resummation effects on the latter quantities by choosing two partly 
rather different input sets for $\Delta u_v $ and $\Delta d_v $. At 
first, we take those of CW \cite{POLSF}, using $x_0 = 0.75$ in eq.~(12) 
of ref.~\cite {POLSF}, which have been used in several theoretical 
investigations~\cite{CO3,BV}.
The small behaviour of this input is relatively flat, $\Delta u_v 
\sim x^{-0.17}$ and $\Delta d_v \sim x^{+0.29}$ at small $x$.
As an example for a more recent parametrization we adopt the
`standard' NLO set of GRSV \cite{GRSV} as an input, using their value
of the scale parameter, $\Lambda_{\overline {\rm MS}} (N_f = 4) = 200 
\MeV$. Here one has $\Delta u_v \sim x^{-0.28}$, $\Delta d_v \sim 
x^{-0.67}$, hence the steepness is similar to that of the unpolarized 
initial distributions above. We have evolved both distribution sets in 
NLO from their respective input scales, $Q^{2} = 10 \mbox{ GeV}^2$ in 
\cite{POLSF} and $Q^{2} = 0.34 \mbox{ GeV}^2$ in \cite{GRSV}, to our
reference scale $Q^2_0 = 4 \GeV^2$. 

Before we derive the quantitative results, we notice that eq.~(\ref
{SOL}) violates the fermion number conservation for the `--' 
non--singlet combinations. Here the conjecture is that the coefficient 
functions $c_{i,l}^{\pm}(x)$ do not contain terms $\propto \ln^{2l} x$ 
in the $\overline{\rm MS}$ scheme. 
For this no proof exists yet, however, we have verified 
this behaviour up to 2--loop order in section~3 for the coefficient 
functions of $xF_3$, $F_2^{NS}$, and $g_1^{NS}$. One should recall
that the main resummation effect comes from that and the next order. 
Under this assumption fermion number conservation has to be restored for
$\GA^-_{x \rightarrow 0}(N,a_s)$. We approach this problem in several 
ways numerically. In a first set of calculations we subtract a 
corresponding term $\propto \delta(1-x)$ from the kernels $K^-$ 
derived from (\ref{eqGAA}), in each order in $a_s$. In $N$--space this
prescription (denoted by `A' below) leads to
\begin{equation}
  \Gamma ^{-}(N,a_s) \rightarrow \Gamma ^{-}(N,a_s) -
  \Gamma ^{-}(1,a_s) \: .
\end{equation}
Another possibility is the restoration of fermion number conservation 
by subleading $1/N$ pole terms. An especially simple choice 
(denoted by `B' in the following) is to modify $\Gamma^-$ according to 
\begin{equation}
  \Gamma ^{-}(N,a_s) \rightarrow \Gamma ^{-}(N,a_s) \cdot (1-N) \: .
\end{equation}
Besides these two prescriptions, which are analogous to the procedure 
in the second reference in~\cite{CCH} with respect to energy--momentum 
conservation in the unpolarized singlet case, we will also show the
results for two other assumptions, namely (`C')
\begin{equation}
 \Gamma ^{-}(N,a_s) \rightarrow \Gamma ^{-}(N,a_s)
 \cdot (1-2N+N^2) \:
\end{equation}
and (`D')
\begin{equation}
 \Gamma ^{-}(N,a_s) \rightarrow \Gamma ^{-}(N,a_s)
 \cdot (1-2N+N^3) \: .
\end{equation}
Clearly, the results of the resummed calculation are only trustworthy, 
and this approach is to be preferred over a fixed order calculation, 
if the difference of the results obtained by all these procedures is 
small.

The corresponding results are presented in figure 2. For the 
relatively flat CW input \cite{POLSF}, the effect is up to about 15\% 
at $x = 10^{-5}$. However, in the kinematical range accessible for
polarized electron and proton scattering at HERA~\cite{JBG1} it again
again amounts to about 1\% or less.
Note, moreover, the wide spread of the results in 
dependence of the employed fermion-number conservation prescription. 
Obviously the resummed contributions do not sufficiently dominate with 
respect to subleading terms at any foreseeable energy. For the steep 
GRSV input \cite{GRSV}, the effect is  of approximately the same
marginal size as in the unpolarized case considered above. In figure~2 
relative corrections are shown. Recall that the absolute values
for $g_1$ obtained in the different parametrizations extrapolating
from the range $x \gsim 10^{-2}$ of the current data down to smaller
$x$ values vary strongly~\cite{JBG1,LAD}.
%
%
\section{Conclusions}
%
We have investigated the resummation of terms of order $\alpha^{l+1}
\ln^{2l} x$, derived in ref.~\cite{KL} for the QCD case, on the
small-$x$ behaviour of non--singlet functions in QED and QCD. The
comparison with the corresponding contributions obtained in the same
order by complete NLO calculations shows the equivalence of both
approaches in this limit up to order $\alpha^2$ in both QED and QCD.
Since the coefficient functions up to two--loop order for the
non--singlet combinations considered contain only terms less singular
in $\ln x$ in the $\overline{\mbox{MS}} $ scheme, the contributions
$\propto a^3 \ln^4 x$ in the three--loop $\overline {\mbox{MS}}$
splitting functions $P^{\pm}(x,a)$ have been predicted on the basis
of this resummation.

A numerical analysis has been performed for the QCD case of
deep-inelastic (polarized) lepton scattering both off unpolarized and
polarized targets.  It turns out that the all--order resummation of the 
terms ${\cal O} (\alpha^{l+1}_s \ln^{2l} x)$ leads only to corrections 
on the level of $1 \%$ in the unpolarized case of $F_2^{\, p} - F_2^{\, 
n}$ even down to extremely small $x$ values, $x = 10^{-15}$. 
The corrections can be larger in the polarized case, up to about 
$15~\%$ at $x \simeq 10^{-5}$, depending on the presently not yet well 
established small-$x$ behaviour of the polarized parton densities. In 
any case, the resummation effects are on the level of 1\% in the 
kinematical range accessible experimentally at present or in the 
foreseeable future.

Presently unknown terms which are suppressed by powers of $\ln x$ in
the splitting functions do contribute in a potentially significant way
to the evolution even at the lowest $x$-values considered. This has 
been demonstrated for the `--' combination $g_1^{\, p} - g_1^{\, n}$ by 
applying several prescriptions to implement fermion number conservation
into the resummed evolution equations. Moreover, the resummation 
corrections are dominated by the first two terms in an $\alpha_s$ 
expansion beyond NLO. All this indicates that fixed-order perturbation 
theory remains the appropriate theoretical framework for the evolution 
of non-singlet structure functions even at very small $x$.

\vspace{5mm}
\noindent
{\bf Acknowledgement}
This work was supported in part by the German Federal Ministry for
Research and Technology under contract No.\ 05 6MU93P.


\newpage
\vspace*{2cm}
\begin{center}

\mbox{\epsfig{file=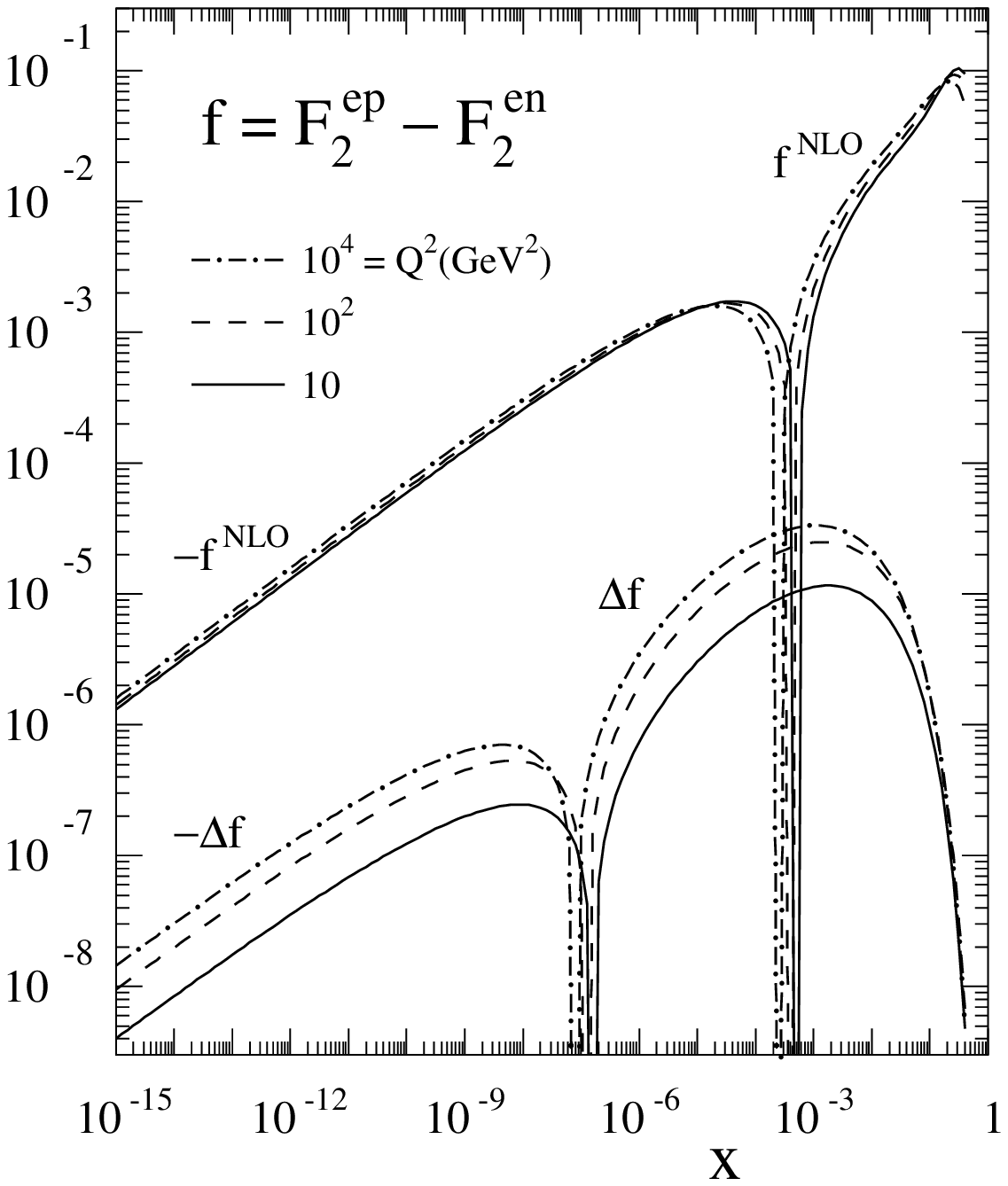,width=13cm}}
\small
\end{center}
{\sf Figure~1:}~The small-$x$ $Q^2$--evolution of the unpolarized 
non--singlet structure function combination $F_2^{\, ep} - F_2^{\, en}$ 
in NLO and the absolute corrections to these results due to the 
resummed kernel derived from ref.~\protect\cite{KL}. The initial
distributions at $ Q_{0}^{2} = 4 \mbox{ GeV}^2 $ have been adopted
from~\protect\cite{MRSA}.

\newpage
\begin{center}

\mbox{\epsfig{file=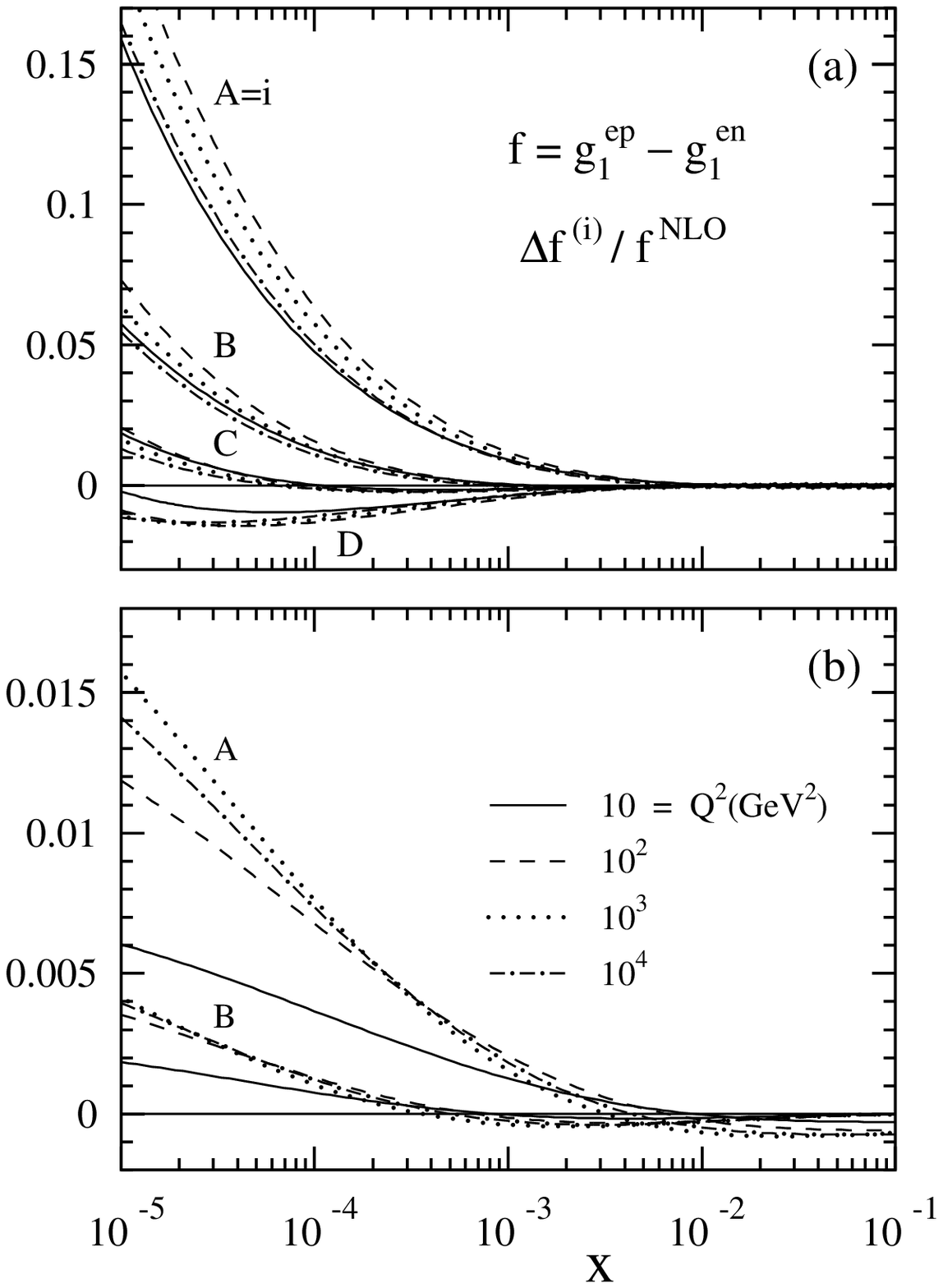,width=13cm}}
\small
\end{center}
{\sf Figure~2:}~The relative corrections to the NLO small-$x$ 
$Q^2$--evolution of the polarized non--singlet structure--function 
difference $g_1^{\, ep} - g_1^{\, en}$ due to the resummed kernel.
`A', `B', `C', and `D' denote the different prescriptions for 
implementing the fermion number conservation discussed in the text.
The initial distribution at $Q_{0}^{2}$ have been taken from 
ref.~\protect\cite{POLSF} in (a) and ref.~\protect\cite{GRSV} in 
(b).

\end{document}